\documentclass[12pt,12pt]{article}
\usepackage{graphicx}
\usepackage{epsfig}
\usepackage{amsmath,amssymb}
\usepackage{slashed} 
\catcode`\@=11
\newdimen\ex@
\font\dozeb=cmmib10 scaled \magstep1
\font\dozesyb=cmbsy10 scaled \magstep1
\font\dezb=cmmib10
\def\bm{\fam9}
\font\mathbf cmbxti10 at 12pt

\def\beq{\begin{equation}}
\def\eeq{\end{equation}}
\def\beqa{\begin{eqnarray}}
\def\eeqa{\end{eqnarray}}
\newcommand{\ba}{\begin{eqnarray}}
\newcommand{\ea}{\end{eqnarray}}
\newcommand\BA{\begin{array}}
\newcommand\EA{\end{array}}

\def\vg{{\bm  g}}
\def\vk{{\bm  k}}

\def\vp{{\bm  p}}
\def\vq{{\bm  q}}

\def\vv{{\bm v}}

\def\vx{{\bm x}}

\def\T{{\bm T}}

\begin{document}

\title{\Large {\bf Velocity operator approach to quantum fluid dynamics\\
\ \\[-20pt]
in a three-dimensional neutron-proton system}
\vspace{-0.5cm}}
\author{Seiya NISHIYAMA\footnotemark[1]~
~and
Jo\~{a}o da PROVID\^{E}NCIA\footnotemark[2]\\
\\[-16pt]
Centro de F\'\i sica,
Departamento de F\'\i sica,\\
\\[-16pt]
Universidade de Coimbra,
P-3004-516 Coimbra, Portugal\footnotemark[2]}

%%%%%% new definition %%%%%%
\def\bm#1{\mbox{\boldmath $#1$}}
\def\bra#1{\langle #1 |}
\def\ket#1{| #1 \rangle }

\date{}

\maketitle

\vspace{-1.0cm}

\footnotetext[1]{~$\!$Corresponding author. 

~~ E-mail address: 
seikoceu@khe.biglobe.ne.jp; nisiyama@teor.fis.uc.pt}
\footnotetext[2]{
$\!$  E-mail address:
providencia@teor.fis.uc.pt}

\vspace{-0.05cm}

%%%%%%%%%%%%%%%%%%%%%%
%                                                                   %
%                              Abstract                        %
%                                                                   %
%%%%%%%%%%%%%%%%%%%%%%

\begin{abstract}
$\!\!\!\!\!\!\!\!\!\!$
In the preceeding paper,
introducing $isospin$-dependent density operators
and defining $\!exact\!$ momenta$\!$
(collective variables),
$\!$we could get an $\!exact\!$ canonically momenta approach
to a one-dimensional $\!$($\!$1D$\!$)$\!$
neutron-proton $\!$($\!$NP$\!$)$\!$ system.
In this paper,
we attempt at a velocity operator approach
to a 3D NP system.
Following Sunakawa,
after introducing momentum density operators,
we define velocity operators,
denoting classical fluid velocities.
We derive a collective 
Hamiltonian in terms of the
collective variables.
\end{abstract}

\vspace{-0.2cm}

 
velocity operator; vortex operator; 
Grassmann variables

\vspace{0.05cm}

PACS Number (s):
21.60.-n, 21.60.Ev

%%%%%%%%%%%%%%%
%                                           %
%    1      Introduction           %
%                                           %       
%%%%%%%%%%%%%%%

\vspace{-0.5cm}

\def\thesection{\arabic{section}}
\section{Introduction}

\vspace{-0.3cm}

~~
To approach elementary excitations in a Fermi system,
Tomonaga gave basic idea in his collective motion theory$\!$ 
\cite{Tomo.50,Emery.79}.
$\!$On the other hand,
it is anticipated that
the Sunakawa's discrete integral equation method
for a Fermi system
\cite{SYN.62}
may also work well for a collective motion problem.
In the preceeding paper
\cite{NishProvi.15}
(referred to as I),
on $isospin$ space $(\!T,\!T_{\!z}\!)$,
introducing
density operators $\rho^{T\!=\!0,T_{\!z}\!=\!0}_{\vk}\!$ and
associated variables $\pi^{0,0}_{\vk}$
and
defining $exact\!$ momenta
$\!\Pi^{0,0}_{\vk}\!\!$
(collective variables),
we could get an $exact\!$ canonically momenta approach
to a one-dimensional ($\!$1D$\!$) neutron-proton ($\!$NP$\!$) system.
In 3D quadrupole nuclear collective motions,
we also have proposed $exact\!$ canonically momenta to collective coordinates
and given
$exact\!$ canonically momenta- and collective coordinate-dependence
of the kinetic part of the Hamiltonian
\cite{NishProvi.14}.
 In this paper,
we attempt at a velocity operator approach
to a 3D NP system.
Following Sunakawa,
after introducing momentum density operators
$\vg^{0,0}_{\vk}$,
we define velocity operators
$\vv^{0,0}_{\vk}$
which denote classical fluid velocities.
We derive a collective 
Hamiltonian in terms of the collective variables
$\vv^{0,0}_{\vk}$ and $\rho^{0,0}_{\vk}$
for irrotational motion.
Its lowest order is diagonalized and
leads us to the Bogoliubov transformation
\cite{Bogoliubov.47}.

In Section 2
first we introduce collective variables $\rho^{0,0}_{\vk}$ 
and associated variables $\pi^{0,0}_{\vk}$ and
give commutation relations between them. 
Next the velocity operator
$\vv^{0,0}_{\vk}$ is defined by a discrete integral equation
and 
commutation relations between the velocity operators are also given. 
In Section 3
the dependence of the original Hamiltonian on
$\vv^{0,0}_{\vk}$ and $\rho^{0,0}_{\vk}$ is determined.
This section is also devoted to a calculation of a constant term
in the collective Hamiltonian.
Finally in Section 4
some discussions and further outlook are given.

%%%%%%%%%%%%%%%%%%%%%%%%%%%
%                                                                                    %
%   2      Collective variable and velocity operator        %
%                                                                                    % 
%%%%%%%%%%%%%%%%%%%%%%%%%%%

\newpage

\def\thesection{\arabic{section}}
\setcounter{equation}{0}
\renewcommand{\theequation}{\arabic{section}.\arabic{equation}}     
\section{Collective variable and velocity operator}

\vspace{-0.4cm}

~~~~In I, we have defined the Fourier component of the density operator
$(\rho(\vx) \!=\! \psi^{\dag}(\vx) \psi(\vx))$
on the $isospin$ space $(T \!=\! 0,T_{z} \!=\! 0)$.
In the 3D system,
they are extended as\\[-18pt]
\ba
\BA{c}
\rho^{0,0}_{\vk}
\!\!\equiv\!\!
{\displaystyle \frac{1}{\sqrt A}} \!
\sum_{\vp, \tau_{z}} \!
a^{\dag}_{\vp+\frac{\vk}{2}, \tau_{z}} \!
a_{\vp-\frac{\vk}{2}, \tau_{z}},~
\rho^{0,0}_{0}
\!\!=\!\!
{\displaystyle \frac{1}{\sqrt{A}}}
\sum_{\vp,\tau_{z}} \!
a^{\dag}_{\vp,\tau_{z}} a_{\vp,\tau_{z}}
\!\!=\!\!
{\displaystyle \frac{N \!+\! Z}{\sqrt{A}}} 
\!\!=\!\!
\sqrt A,
\EA \!\!
\label{FcomponentDensityOp}
\ea\\[-14pt]
where we $\!$have $\!$used the anti commutation relation (CR)s among
$a_{\vk, \tau_{z}}$'s and $a^{\dag}_{\vk, \tau_{z}}$'s
given by\\[-18pt]
\ba
\BA{c}
\{ \! a_{\vk, \tau_{z}}, a^{\dag}_{\vk^{\prime}, \tau^{\prime}_{z}} \! \}
\!=\!
\delta_{\vk, \vk^{\prime}}
\delta_{\tau_{z}, \tau^{\prime}_{z}} ,~
\{ \! a_{\vk, \tau_{z}}, a_{\vk^{\prime}, \tau^{\prime}_{z}} \! \}
\!=\!
\{ \! a^{\dag}_{\vk, \tau_{z}}, a^{\dag}_{\vk^{\prime}, \tau^{\prime}_{z}} \! \}
\!=\!
0 ,
\EA
\label{ACR}
\ea\\[-16pt]
where
$\!N\!$ and $\!Z\!$ mean the number of
neutron$(\tau_{z} \!=\! \frac{1}{2})\!$ and
proton$(\tau_{z} \!=\!\!-\frac{1}{2})\!\!$
\cite{Lipkin.65,RoweWood.2010}.$\!$
We here consider a spin-less fermion.
The Hamiltonian $H$
in a 3D box $\Omega~(\!=\! L^3)$
is given by
Eq. (2.10) in I as\\[-20pt]
\ba
\BA{c}
H
\!=\!
T \!+\! V
\!=\!
\sum_{\vk,\tau_{z}} \!
{\displaystyle \frac{\hbar^{2}\vk^{2}}{2m}}
a^{\dag}_{\vk \tau_{z}}
a_{\vk \tau_{z}}
\!\!-\!
{\displaystyle \frac{A}{8\Omega}} \!
\sum_{\vk}
\nu_{T=0}(\vk)
\rho^{0, 0}_{\vk}
\rho^{0, 0}_{-\vk}
\!-\!
{\displaystyle \frac{A}{4\Omega}} \!
\sum_{\vk}
\nu_{T=0}(\vk) ,
\EA
\label{ExpressionforHamiltonian}
\ea\\[-14pt]
where
$\nu_{T=0}$, 
denoted simply as $\nu$, 
is a scaler function of $\vk$
which specifies the interaction.

Following Tomonaga
\cite{Tomo.50},
first we introduce associated collective momenta
$\pi_{\vk}^{0,0}$
defined by\\[-16pt]
\ba
{\pi_{\vk}^{0,0}}
\!\equiv\!
{\displaystyle \frac{m} {\vk^{2}}}
\dot{ \left(\rho_{-\vk}^{0,0}\right)}
\!=\!
{\displaystyle \frac{m} {\vk^{2}}}
{\displaystyle \frac{i} {\hbar}}
[H, \rho^{0,0}_{-\vk} ]
\!=\!
{\displaystyle \frac{m} { \vk^{2}}}
{\displaystyle \frac{i} {\hbar}}
[T, \rho^{0,0}_{-\vk} ]  ,
~(\vk \!\ne\! 0) ~,
\label{piplusminusk} 
\ea\\[-14pt]
where the upper symbol dot $\!\cdot\!$ means a time derivative.
Calculating the commutator
(\ref{piplusminusk}),
we obtain the explicit 
expression for the associated collective variables
$\pi_{\vk}^{0,0}$
given as\\[-18pt]
\ba
\BA{c}
{\pi_{\vk}^{0,0}}
\!=\!
-{\displaystyle \frac{i \hbar}{\sqrt A \vk \!\cdot\! \vk}}
\sum_{\vp, \tau_{z}} \!
\vk \!\cdot\! \vp
a^{\dag}_{\vp-\frac{\vk}{2}, \tau_{z}} \!
a_{\vp+\frac{\vk}{2}, \tau_{z}} \! ,
(\mbox{The symbol $\!\cdot\!$ denotes an inner product}.) .
\EA
\label{piplusminusk2}  
\ea\\[-14pt]
The CRs among
the density $\rho^{0,0}_{\!\vk}\!$ and
the associated $\pi^{0,0}_{\!\vk}\!$ operators,
however,
become as follows:\\[-16pt]
\ba
\BA{ll}
\left[ \rho^{0,0}_{\vk}, \rho^{0,0}_{\vk^{\prime}} \right]
\!=\!
0 , ~~
[\pi^{0,0}_{\vk} ,\rho^{0,0}_{\vk'}]
\!=\!
-
{\displaystyle \frac{i\hbar }{\sqrt{A}}}
{\displaystyle \frac{\vk\!\cdot\!\vk' }{\vk\!\cdot\!\vk}}
\rho^{0,0}_{\vk'-\vk} ,~~
[\pi^{0,0}_{\vk} ,\pi^{0,0}_{\vk'}]
\!=\!
- {\displaystyle \frac{i\hbar }{\sqrt{A} }}
{\displaystyle \frac{\vk\!\cdot\!\vk \!-\! \vk'\!\cdot\!\vk' }{\vk\!\cdot\!\vk'}}
\pi^{0,0}_{\vk+\vk'} .
\EA
\label{CRpirho}  
\ea\\[-12pt]
From now on,
following Sunakawa
\cite{SYN2.62}, 
we introduce a momentum density operator given by\\[-18pt]
\ba
\BA{c}
\vg^{0,0}_{\vk}
\!\!\equiv\!\!
{\displaystyle \frac{\hbar}{\sqrt A}} \!
\sum_{\vp, \tau_{z}} \!
\vp 
a^{\dag}_{\vp-\frac{\vk}{2}, \tau_{z}} \!
a_{\vp+\frac{\vk}{2}, \tau_{z}}
\!\!=
i \vk \pi^{0,0}_{\vk}, ~~
\vg^{0,0}_{0}
\!\!=\!
{\displaystyle \frac{\hbar}{\sqrt{A}}} \!
\sum_{\vp,\tau_{z}}
\vp
a^{\dag}_{\vp,\tau_{z}} a_{\vp,\tau_{z}}
\!=\!
0.
\EA
\label{FcomponentMomentumDensityOp}
\ea\\[-14pt]
In the $\vg^{0,0}_{0}\!$,
we assume zero-eigenvalue of the total momentum of our system.
The CRs among $\rho^{0,0}_{\vk}$  
and vector $\vg^{0,0}_{\vk}~
(=(g_{\vk}^{0,0(1)},g_{\vk}^{0,0(2)},g_{\vk}^{0,0(3)}))$
with vector ${\vk}~(=(k_1,k_2,k_3))$
are obtained as\\[-12pt]
\ba
\BA{l}
[\vg^{0,0}_{\vk} ,\rho^{0,0}_{\vk'}]
\!=\!
{\displaystyle \frac{\hbar \vk'}{\sqrt{A}}}
\rho^{0,0}_{\vk'-\vk} ,~~
[g^{0,0(i)}_{\vk} ,g^{0,0(j)}_{\vk'}]
\!=\!
{\displaystyle \frac{\hbar }{\sqrt{A}}} \!
\left( \!
g^{0,0(i)}_{\vk+\vk'}k_j
-
g^{0,0(j)}_{\vk+\vk'}k'_i \!
\right) .
\EA
\label{CRgrho}  
\ea\\[-28pt]

Following Sunakawa
\cite{SYN2.62}, 
we define the modified momentum density operator
$\vv^{0,0}_{\vk}$ by\\[-18pt]
\ba
\BA{c}
\vv^{0,0}_{\vk}
\!\equiv
\vg^{0,0}_{\vk} 
\!-\!
{\displaystyle \frac{1}{ \sqrt {A}}}
\sum_{\vp \ne \vk} 
\rho^{0,0}_{\vp-\vk} \vv^{0,0}_{\vp}~ 
(\vk \!\ne\!  0) ,~~
\vv^{0,0}_0
\!=\!
0 .
\EA 
\label{modifiedv}  
\ea\\[-16pt]
Along the same way as the one in I and the Sunakawa's$\!$
\cite{SYN.62},
we can prove two important CRs\\[-18pt] 
\ba
[\vv^{0,0}_{\!\vk}, \rho^{0,0}_{\!\vk'}]
\!=\! 
\hbar \vk' {\delta }_{\vk,\vk'} ,
\label{modifiedCRs}  
\ea\\[-30pt]
\ba
\!\!\!\!\!\!
\BA{c}
[v^{0,0(i)}_{\vk} \! , v^{0,0(j)}_{\vk'}]
\!\approx\!
-
{\displaystyle \frac{\hbar}{A} } \!\!
\sum_{\vp\mbox{\scriptsize{all}}} \! \rho^{0,0}_{\!\vp-\vk-\vk'} \!
\left( \! p_i v^{0,0(j)}_{\!\vp} \!-\! p_j v^{0,0(i)}_{\!\vp} \! \right)
\!+\!
{\displaystyle \frac{1}{A} } \!\!
\sum'_{\vp,\vq} \! \rho^{0,0}_{\!\vp-\vk}\rho^{0,0}_{\!\vq-\vk'}
[v^{0,0(i)}_{\!\vp} \! , v^{0,0(j)}_{\!\vq}]  .
\EA 
\label{CRvv}  
\ea\\[-14pt]
The symbol $\!\sum'\!$ means that
the term
$\vp\!\!=\!\!\vk$ and $\vq\!\!=\!\!\vk'$$\!$
at the same time
should be omitted.
The CRs
(\ref{modifiedCRs})
and
(\ref{CRvv})
are quite the same as the Sunakawa's
\cite{SYN2.62}.
As pointed out by him,
Fourier transforms of the operators
$\rho^{0,0}_{\vk}$ and $\vv^{0,0}_{\vk}$
and the CRs among them
are identical with those found by Landau
\cite{Landau.71}
for the fluid dynamical density operator and the velocity operator.
Then, it turns out that the quantum mechanical operator
$\!\vv(\vx\!)$,
which satisfies the famous CR\\[-22pt]
\ba
\BA{c}
[v^{0,0(i)}(\vx) , v^{0,0(j)}(\vx')]
\!=\!
{\displaystyle \frac{i\hbar}{m}} 
\delta(\vx-\vx')
\rho^{0,0}(\vx)^{-1}(\mbox{rot}\vv^{0,0}(\vx))^{(k)}~
(i,~j,~k)~\mbox{cyclic},
\EA 
\label{CRvivj}  
\ea\\[-14pt]
corresponds to the fluid dynamical velocity.
We also have another well-known CR\\[-18pt]
\ba
\BA{c}
[\vv^{0,0}(\vx), \rho^{0,0}(\vx')]
\!=\!
-
{\displaystyle \frac{i\hbar}{m}}
\nabla_{\!x}
\delta(\vx-\vx').
\EA 
\label{CRvrho}  
\ea

%%%%%%%%%%%%%%%%%%%%%%%%%%%%%%%%
%                                                                                                    %
%      3      v_k- and rho_k-dependence of the Hamiltonian          %
%                                                                                                    % 
%%%%%%%%%%%%%%%%%%%%%%%%%%%%%%%%

\newpage

\def\thesection{\arabic{section}}
\setcounter{equation}{0}
\renewcommand{\theequation}{\arabic{section}.\arabic{equation}}     
\section{$\vv_{\vk}$- and $\rho_{\vk}$-dependence of the Hamiltonian}

~~~~We derive here a collective 
Hamiltonian in terms of the
$\vv^{0,0}_{\vk}$ and $\rho^{0,0}_{\vk}$.
Following Sunakawa, we expand
the kinetic operator $T$
in a power series of the velocity operator
$\vv^{0,0}_{\vk}$ 
as follows:\\[-20pt] 
\ba
\!\!
\BA{c}
T
\!\!=\!\! 
{T}_{0}(\rho)
\!\!+\!\! 
\sum_{\vp \ne 0} \!
{\T}_{\!1} \! (\rho ; \vp  ) \!\cdot\! \vv^{0,0}_{\vp}
\!\!+\!\!
\sum_{\vp \ne 0, \vq \ne 0} \!
{T}_{\!2} (\rho ; \vp,\vq )
\vv^{0,0}_{\vp} \!\cdot\! \vv^{0,0}_{\vq}
\!\!+\! 
\cdot \cdot \cdot  ,~
{T}_{\!2} (\rho ; \vp,\vq )
\!\!=\!\! 
{T}_{\!2} ( \rho ; \vq,\vp ) ,
\EA
\label{Texpansion}  
\ea\\[-20pt]
in which 
$\!{T_{n} (n \!\neq\! 0)}\!$ are unknown expansion coefficients. 
In order to determine their explicit expressions, we take the commutators 
between $T$ and $\rho^{0,0}_{\vk}$
as follows:\\[-16pt]
\ba
\!\!
\left.
\BA{ll}
&
[T , \rho^{0,0}_{\vk}]
\!=\! 
\hbar
{\T}_{1} (\rho ; \vk  ) \!\cdot\! \vk
\!+\!
2 \hbar \!
\sum_{\vp \ne 0} \!
{T}_{2} (\rho ; \vp,\vk )
\vv^{0,0}_{\vp} \!\cdot\! \vk
\!+\! 
\cdots  ,~\\
\\[-6pt]
&
[[T, \rho^{0,0}_{\vk}] , \rho^{0,0}_{\vk'}]
\!=\!
2 \hbar^2 
{T}_{2} ( \rho ; \vk',\vk ) \vk' \!\cdot\! \vk
\!+\! 
\cdots  .
\EA \!\!
\right\}
\label{Texpansioncommrho}  
\ea\\[-14pt]
On the other hand,
from  
(\ref{FcomponentMomentumDensityOp}) and (\ref{modifiedv}),
we can calculate the commutators 
between ${T}$ and ${\rho^{0,0}_{\vk}}$ by using the relations
(\ref{piplusminusk})
and
(\ref{FcomponentMomentumDensityOp}) 
as follows:\\[-20pt]
\ba
\BA{c}
[T , \rho^{0,0}_{\vk}]
\!=\!
{\displaystyle \frac{\hbar }{m}}
\vk
\!\cdot\!
\vg^{0,0}_{-\vk}
\!=\!
{\displaystyle \frac{\hbar }{m}}
\vk
\!\cdot\!
\vv^{0,0}_{-\vk}
\!+\!
{\displaystyle {\frac{\hbar}{m \! \sqrt {\!A}}}} \!\!
\sum_{\vp \ne -\vk} \!
\rho^{0,0}_{\vp+\vk}
\vk
\!\cdot\!
\vv^{0,0}_{\vp}  ,
\EA
\label{CRTrho}  
\ea\\[-16pt]
and using the relation 
(\ref{modifiedCRs})
successively, we can easily obtain the following commutators:\\[-22pt] 
\ba
[[T, \rho^{0,0}_{\vk}] , \rho^{0,0}_{\vk'}]
\!=\!\!\!
\BA{c}
-{\displaystyle \frac{ \hbar^{2} {\vk}^{2}}{m}}\delta_{\vk',-\vk }~,~
{\displaystyle \frac{{\hbar}^{2}}{m \! \sqrt {\!A}}}
\vk
\!\cdot\!
\vk'
\rho^{0,0}_{\vk+\vk'}~,~\vk'  \!\ne\!  -\vk ,
\EA
\label{CRTrhorho}
\ea\\[-36pt]
\ba
[[[T , \rho^{0,0}_{\vk} ] , \rho^{0,0}_{\vk'}] , \rho^{0,0}_{\vk''}]
\!=\!
0 ,~
\!\cdots\! .
\label{CRTrhorhorho} 
\ea\\[-20pt]
Comparing the above results with the commutators
(\ref{Texpansioncommrho}),
we can determine the coefficients 
${T_{n} (n \!\neq\! 0)}$. Then we can express the kinetic part ${T}$ in terms 
of the 
$\rho^{0,0}_{\vk}$ and $\vv^{0,0}_{\vk}$ as follows:\\[-22pt]
\ba
\BA{c}
T
\!=\!
{T}_{0}(\rho)
\!+\!
{\displaystyle \frac{1}{2m}} \!
\sum_{\vk \ne 0} \!
\vv^{0,0}_{\vk}
\!\cdot\!
\vv^{0,0}_{-\vk}
\!+\! 
{\displaystyle \frac{1}{2m \! \sqrt {\!A}}} \!
\sum_{\vp+\vq \ne 0} \!
\rho^{0,0}_{\vp+\vq}
\vv^{0,0}_{\vp}
\!\cdot\!
\vv^{0,0}_{\vq} ,~
(\vv^{0,0}_0 \!=\! 0) .
\EA
\label{exactTPi}  
\ea\\[-20pt]
Up to the present stage, all the expressions 
have been derived without any approximation. 

Our remaining task is to determine the term
$T_{0}(\rho)$ in 
(\ref{Texpansion})
which depends only on $\rho^{0,0}_{\vk}$. 
For this aim, we also expand it in a power series of the collective 
coordinates $\rho^{0,0}_{\vk}$ in the form\\[-22pt] 
\ba
\BA{c}
{T}_{0} (\rho)
\!=\! 
{C}_{0}
\!+\!
\sum_{\vp \ne 0} \!
{C}_{1}(\vp) \rho^{0,0}_{\vp}
\!+\!
\sum_{\vp \ne 0, \vq \ne 0} \!
{C}_{2} (\vp,\vq)
\rho^{0,0}_{\vp} \! \rho^{0,0}_{\vq} 
\!+\! 
\cdot \cdot \cdot  ,~
{C}_{2} (\vp,\vq ) \!=\!  {C}_{2} (\vq,\vp ) ,
\EA
\label{T0rho}  
\ea\\[-22pt]
and to get expansion coefficients,
we take the commutators 
between ${T}$ and ${\vv^{0,0}_{\vk}}$
as follows:\\[-20pt]
\ba
\!\!
\left.
\BA{ll}
&
[v^{(i)}_{\vk}  , {T}_{0}(\rho)]
\!=\! 
\hbar k_i
{C}_{1} (\vk  ) 
\!+\!
2 \hbar k_i \!
\sum_{\vp \ne 0} 
{C}_{2} (\vp ; \vk )
\rho^{0,0}_{\vp}
\!+\! 
\cdots  ,~\\
\\[-6pt]
&
[v^{(j)}_{\vk'} , [v^{(i)}_{\vk}  , {T}_{0}(\rho)]]
\!=\!
2 \hbar^2 k_i k'_j
{C}_{2} ( \vk' ; \vk ) 
\!+\!
6 \hbar^2 k_i k'_j \!
\sum_{\vp \ne 0} 
{C}_{3} (\vp ; \vk' ; \vk ) 
\rho^{0,0}_{\vp}
\!+\!
\cdots  .
\EA \!\!
\right\}
\label{Texpansioncommvelo}  
\ea\\[-14pt]
We here restrict our Hilbert space to subspace
in which eigenvalue of the {\bf vortex operator}
satisfies
$\mbox{rot}\vv^{0,0}(\!\vx\!)| \!\! >=\!0, \!$
i.e.,
$\!
[v^{0,0(i)}_{\vk} \! , v^{0,0(j)}_{\vk'}]
\!=\!
0
$.
From 
(\ref{modifiedv}),
we have a discrete integral equation\\[-20pt]
\ba
\BA{c}
[v^{(i)}_{\vk}  , {T}_{0}(\rho)]
\!=\!
{f}^{(i)}(\rho ; \vk)
\!-\!
{\displaystyle \frac{1}{\sqrt {\!A}}}
\sum_{\vp \ne 0, \vk}
\rho^{0,0}_{\vp-\vk}
[v^{(i)}_{\vp}  , {T}_{0}(\rho)]  .
\EA
\label{CRPiT0}  
\ea\\[-18pt]
With the aid of
(\ref{exactTPi})
and using two CRs of
(\ref{CRgrho}),
the inhomogeneous term ${f}^{(i)}(\rho ; \vk)$ becomes\\[-20pt]
\ba
\!\!\!\!\!\!\!\!
\BA{ll}
&f^{(i)}\!(\rho ; \vk)
\!\equiv\! 
[g^{0,0(i)}_{\vk} , {T}_{\!0}(\rho)]
\!=\!\!
\left[ 
g^{0,0(i)}_{\vk} ,
T 
\!-\!
{\displaystyle \frac{1}{2m}} \!\!
\sum_{\vk \ne 0 } \!\!  
\vv^{0,0}_{\vk} \!\cdot\! \vv^{0,0}_{-\vk} 
\!-\! 
{\displaystyle \frac{1}{ 2m \! \sqrt {\!A}}} \!\!
\sum_{\vp+\vq \ne 0} \!
\rho^{0,0}_{\vp+\vq}
\vv^{0,0}_{\vp} \!\cdot\! \vv^{0,0}_{\vq} 
\right] \\
\\[-10pt]
\!\!\!\!
&=\!
[g^{0,0(i)}_{\vk} , T]
\!-\!
{\displaystyle \frac{\hbar }{m \! A}} \!\!
\sum_{\!\vp, \vq~\!\mbox{\scriptsize all}} \!
p_i
\rho^{0,0}_{\!\vp+\vq-\vk} 
\vv^{0,0}_{\vp} \!\cdot\! \vv^{0,0}_{\vq} 
\!\!-\!\!
{\displaystyle \frac{\hbar }{m \! A}} \!\!
\sum_{\!\vp, \vq~\!\mbox{\scriptsize all}} 
\rho^{0,0}_{\!\underline{\vp}+\vq} \!\!
\sum_j \!\!
\left( \!\!
g^{0,0(i)}_{\vk+\underline{\vp}} k_j 
\!\!-\!\!
g^{0,0(j)}_{\vk+\underline{\vp}} p_i  \!\!
\right) \!\!
\vv^{0,0(j)}_{\vq} \\
\\[-8pt]
\!\!\!\!
&=\!
[g^{0,0(i)}_{\vk} , T]
\!-\!
{\displaystyle \frac{\hbar }{m \! A}} \!\!
\sum_{\!\vp, \vq~\!\mbox{\scriptsize all}} \!
\rho^{0,0}_{\!\vp+\vq-\vk} \!
(\vk \!\cdot\! \vv^{0,0}_{\vp}) \vv^{0,0(i)}_{\vq} \! ,
(\mbox{due to change of the variable}~\underline{\vp}
~\mbox{to}~\vp\!\!-\!\!\vk)  .
\EA
\label{fkrho}  
\ea\\[-8pt]
To obtain the explicit formula for
${f}^{(i)}(\rho ; \vk)$,
first we calculate the CR between
$g^{0,0(i)}_{\vk}$ and ${T}$
as \\[-20pt]
\ba
\!\!\!\!\!\!
[g^{0,0(i)}_{\vk} \! , T]
\!\!=\!\!
{\displaystyle \frac{ \hbar^{3}}{2m \! \sqrt {\!\!A}}} \!
\sum_{\vp~\!\!\mbox{\scriptsize all}, \tau_{\!z}} \!\!\!
p_i \!\!
\left\{ \!\!\!
\left( \!\! \vp \!+\! {\displaystyle \frac{\vk}{2}} \!\! \right)^{\!\!\!2}
\!\!-\!\!
\left( \!\! \vp \!-\! {\displaystyle \frac{\vk}{2}} \!\! \right)^{\!\!\!2} \!\!
\right\} \!\!
a^{~\!\!\dag}_{\!\vp-\frac{\vk}{2}, \tau_{\!z}} \!\!
a_{\vp+\frac{\vk}{2}, \tau_{\!z}}
\!\!\!=\!\!
{\displaystyle \frac{ \hbar^{3}}{ m \! \sqrt {\!\!A}}} \!
\sum_{\vp~\!\!\mbox{\scriptsize all}, \tau_{\!z}} \!\!\!
p_i (\vp \!\cdot\! \vk) \!
a^{~\!\!\dag}_{\!\vp-\frac{\vk}{2}, \tau_{\!z}} \!\!
a_{\vp+\frac{\vk}{2}, \tau_{\!z}} \! .
\label{commupiT}  
\ea\\[-26pt] 

From now on, we make an approximation for
$\rho^{0,0}_{0} \!\!=\!\! \sqrt {\!A},\vv^{0,0}_{\vk}$ and
$\rho^{0,0}_{\vk}$
as\\[-20pt]
\ba
\BA{c}
\vv^{0,0}_{\vk}
\!\cong\!
{\displaystyle \frac{\hbar \vk}{2}} \!
\sum_{\tau_{z}} \!\!
\left( 
\overline{\theta}a_{\vk, \tau_{z}}
\!-\!
a^{\dag}_{-\vk, \tau_{z}}\theta 
\right) ,~
\rho^{0,0}_{\vk}
\!\cong\!
\sum_{\tau_{z}} \!\!
\left( 
\overline{\theta}a_{-\vk, \tau_{z}}
\!+\!
a^{\dag}_{\vk, \tau_{z}}\theta 
\right) . 
\EA
\label{approxpirho}  
\ea\\[-16pt]
The operators
$a_{0, \tau_{z}}$ and  $a^{\dag}_{0, \tau_{z}}$
are regarded
as $c$-numbers but with the Grassmann variables
which play crucial roles to compute the inhomogeneous term
${f}^{(i)}(\rho ; \vk)$
(\ref{CRPiT0}),
as shown in I.
The explicit forms of the operators are simply given as\\[-18pt]
\ba
a_{0, \tau_{z}}
\!\cong\!
\sqrt{\!A} ~\! \theta ,~
a^{\dag}_{0, \tau_{z}}
\!\cong\!
\sqrt{\!A} ~\! \overline{\theta} ,
\label{approxa0}  
\ea\\[-22pt]
where the 
$\theta$ and $\overline{\theta}$
are the Grassmann variables and anti-commute with
$a_{\vk, \tau_{z}}$ and $a^{\dag}_{\vk, \tau_{z}}$
\cite{Berezin.66,Casalbuoni.76a,Casalbuoni.76b}.
Then the second term in the last line of 
(\ref{fkrho})
is approximately computed as\\[-22pt]
\ba
\BA{c}
-
{\displaystyle \frac{\hbar }{m \! A}} \!
\sum_{\vp, \vq~\!\mbox{\scriptsize all}}
\rho^{0,0}_{\vp+\vq-\vk} 
(\vk \!\cdot\! \vv^{0,0}_{\vp}) \vv^{0,0(i)}_{\vq} \\
\\[-10pt]
\!=\!
-
{\displaystyle \frac{\hbar^3}{4 m \! \sqrt{\!A}}} \!
\sum_{\vp~\!\mbox{\scriptsize all}} 
(\vp \!\cdot\! \vk) (k_i \!-\! p_i) \!
\sum_{\tau_{z}} \!
\left( 
\overline{\theta}a_{\vp, \tau_{z}} 
\!-\!
a^{\dag}_{-\vp, \tau_{z}}\theta 
\right) \!\!
\sum_{\tau'_{z}} \!
\left( 
\overline{\theta}a_{\vk-\vp, \tau'_{z}} 
\!-\!
a^{\dag}_{-(\vk-\vp), \tau'_{z}}\theta 
\right) \\
\\[-10pt]
\!=\!
-
{\displaystyle \frac{\hbar^3}{4 m \! \sqrt{\!A}}} \!
\sum_{\vp~\!\mbox{\scriptsize all}} 
(\vp \!\cdot\! \vk) (k_i \!-\! p_i) \!
\left( 
\rho^{0,0}_{-\vp} 
\!-\!
2 \! \sum_{\tau_{z}} \! a^{\dag}_{-\vp, \tau_{z}}\theta 
\right) \!\!
\left( 
\rho^{0,0}_{-(\vk-\vp)} 
\!-\!
2 \! \sum_{\tau'_{z}} \! a^{\dag}_{-(\vk-\vp), \tau'_{z}}\theta 
\right) \\
\\[-10pt]
\!=\!
-
{\displaystyle \frac{\hbar^3}{4 m \! \sqrt{\!A}}} \!\!
\sum_{\vp~\!\mbox{\scriptsize all}} 
(\vp \!\cdot\! \vk) (k_i \!\!-\!\! p_i) \!
\left\{ \!
\rho^{0,0}_{\!-\vp}
\rho^{0,0}_{\!\vp \!-\! \vk\!} 
\!\!-\!
2 \!\!
\sum_{\tau_{z},\tau'_{z}} \!\!
\left( \!
\theta \overline{\theta} 
a^{\dag}_{-\vp, \tau_{z}} \!
a_{\vk \!-\! \vp, \tau'_{z}}
\!\!+\!\!
\overline{\theta} \theta 
a_{\vp, \tau_{z}} \!
a^{\dag}_{-(\vk \!-\! \vp), \tau'_{z}} \!
\right) \!
\right\} \\
\\[-12pt]
\!\!\!=\!
-
{\displaystyle \frac{\hbar^3}{4 m \! \sqrt{\!A}}} \!
\sum_{\vp~\!\mbox{\scriptsize all}} 
(\vp \!\cdot\! \vk) (k_i \!-\! p_i) 
\rho^{0,0}_{-\vp}
\rho^{0,0}_{\!\vp \!-\! \vk\!} \\
\\[-8pt]
-
\theta \overline{\theta}
{\displaystyle \frac{\hbar^3}{2 m \! \sqrt{\!A}}} \!\!
\sum_{\vp~\!\mbox{\scriptsize all}} \!
\sum_{\tau_{z},\tau'_{z}} \!
\left\{ \!
\left( \! \vp \!\cdot\! \vk \!-\! {\displaystyle \frac{\vk^2}{2}} \! \right) \!\!
\left( \! p_i \!+\!{\displaystyle \frac{ k_i}{2}} \! \right)  
\!+\!
\left( \! \vp \!\cdot\! \vk \!+\! {\displaystyle \frac{\vk^2}{2}} \! \right) \!\!
\left( \! p_i \!-\!{\displaystyle \frac{ k_i}{2}} \! \right) \!
\right\} \!
a^{\dag}_{\vp \!-\! \frac{\vk}{2}, \tau_{z}} \!\!
a_{\vp \!+\! \frac{\vk}{2}, \tau'_{z}} \\
\\[-10pt] 
\!\!\!=\!
-
{\displaystyle \frac{\hbar^3}{4 m \! \sqrt{\!A}}} \!\!
\sum_{\vp~\!\mbox{\scriptsize all}} 
(\vp \!\cdot\! \vk) (k_i \!\!-\!\! p_i) 
\rho^{0,0}_{-\vp}
\rho^{0,0}_{\!\vp \!-\! \vk\!}
\!-\!
\theta \overline{\theta} 
{\displaystyle \frac{\hbar^3}{m \! \sqrt{\!A}}} \!\!
\sum_{\vp~\!\mbox{\scriptsize all}} \!\!
\left( \!
\vp \!\cdot\! \vk p_i
\!\!-\!\! 
{\displaystyle \frac{\vk^2}{4}} k_i \!\!
\right) \!\!
\sum_{\tau_{z},\tau'_{z}} \!
a^{\dag}_{\vp \!-\! \frac{\vk}{2}, \tau_{z}} \!\!
a_{\vp \!+\! \frac{\vk}{2}, \tau'_{z}} \\
\\[-12pt] 
\!\!\!=\!
-
{\displaystyle \frac{\hbar^3}{4 m \! \sqrt{\!A}}} \!\!
\sum_{\vp \mbox{\scriptsize all}} 
(\vp \!\cdot\! \vk) (k_i \!\!-\!\! p_i) 
\rho^{0,0}_{-\vp}
\rho^{0,0}_{\!\vp \!-\! \vk\!}
\!-\!
\theta \overline{\theta} 
{\displaystyle \frac{\hbar^3}{m }} \!
\left( \!
{\displaystyle \frac{\vk}{2}} \!\cdot\! \vk
{\displaystyle \frac{k_i}{2}}
\!-\! 
{\displaystyle \frac{\vk^2}{4}} k_i \!
\right) \!\!
\left(
\overline{\theta}
a_{\vk, \tau'_{z}}
\!+\!
a^{\dag}_{-\vk}
\theta
\right) , \\
\EA
\label{approxrhokvv}  
\ea\\[-12pt]
the last term of which
is obtained by extracting the terms with
$\vp \!=\! {\displaystyle \frac{\vk}{2}}$
or
$\vp \!=\! -{\displaystyle \frac{\vk}{2}}$
in the last term in the second line from the bottom
and it evidently vanishes.

Substituting the resultant formula of
(\ref{commupiT}),
$
[g^{0,0(i)}_{\vk}, T]
\!\!=\!\!
{\displaystyle {\frac{{\hbar }^{3} k_i {\vk}^{2}}{4m}}} 
\rho^{0,0}_{-\vk}
$,
and
the calculated result
(\ref{approxrhokvv})
into
${{f}^{(i)}(\rho ; \vk)}$, i.e.,
(\ref{fkrho}),
we get an approximate formula for the
${{f}^{(i)}(\rho ; \vk)}$ up to the order of
$\frac{1}{\sqrt {\!A}}$
in the following form:\\[-20pt]
\ba
\!\!\!\!
\BA{c}
{f}^{(i)}(\rho ; \vk)
\!=\!
{\displaystyle {\frac{{\hbar }^{3} k_i {\vk}^{2}}{4m}}} \!
\rho^{0,0}_{-\vk}
\!-\!
{\displaystyle \frac{\hbar^3}{4 m \! \sqrt{\!A}}}
\sum_{\vp \ne \vk} 
(\vp \!\cdot\! \vk ) (k_i - p_i)
\rho^{0,0}_{-\vp}
\rho^{0,0}_{\vp \!-\! \vk} ~.
\EA
\label{approxfkrho}  
\ea\\[-14pt]
Further substituting 
(\ref{approxfkrho}) into 
(\ref{CRPiT0})
and bringing the next leading term, 
we can rewrite the R.H.S. of the discrete integral equation
(\ref{CRPiT0}) as\\[-20pt]
\ba
\!\!\!\!
\BA{ll} 
&[v^{0,0(i)}_{\vk} , {T}_{0} (\rho)]
\!=\! 
{\displaystyle {\frac{{\hbar }^{3} k_i {\vk}^{2}}{4m}}} \!
\rho^{0,0}_{-\vk}
\!-\!
{\displaystyle \frac{\hbar^3 }{4 m \! \sqrt{\!A}}} \!\!
\sum_{\vp \ne \vk} \!
\left\{
(\vp \!\cdot\! \vk ) (k_i \!-\! p_i)
\!+\!
p_i \vp^2
\right\} \!
\rho^{0,0}_{-\vp}
\rho^{0,0}_{\!\vp \!-\! \vk\!} \\
\\[-12pt] 
&\approx\! 
{\displaystyle {\frac{{\hbar }^{3} k_i {\vk}^{2}}{4m}}} \!
\rho^{0,0}_{-\vk}
\!-\!
{\displaystyle \frac{\hbar^3 k_i }{8 m \! \sqrt{\!A}}} \!\!
\sum_{\vp \ne \vk} \!
\vp \!\cdot\! (\vp \!+\! \vk)
\rho^{0,0}_{-\vp}
\rho^{0,0}_{\vp \!-\! \vk} ~,
\left( \! \mbox{under the assumption of}~p_i \!=\! {\displaystyle \frac{k_i}{2}} \! \right) .
\EA
\label{CRPiT02}  
\ea
From
(\ref{CRPiT02}) and the CRs
(\ref{CRpirho})
and 
(\ref{modifiedCRs}),
we get the following commutation relations:\\[-20pt] 
\ba
\!\!\!\!\!\!\!\!\!\!\!\!
\left.
\BA{ll}
&
[v^{0,0(j)}_{\vk'} \! , [v^{0,0(i)}_{\vk} \! , {T}_{0} (\rho)]]
\!=\! 
- 
{\displaystyle {\frac{{\hbar }^{4} k_i k_j {\vk}^{2}}{4m}}}
{\delta }_{\vk', -\vk} 
\!-\! 
{\displaystyle {\frac{{\hbar }^{4} k_i k'_j}{4 m \! \sqrt {\!A}}}} \!
\left( \!
\vk^2
\!+\!
\vk \!\cdot\! \vk'
\!+\!
\vk'^2 \!
\right) \!
\rho^{0,0}_{-\vk-\vk'} , \\
\\[-12pt]
&
[v^{0,0(k)}_{\vk''} \! , [v^{0,0(j)}_{\vk'} \! , [v^{0,0(i)}_{\vk} \! , {T}_{0}(\rho)]]]
\!=\!
-  
{\displaystyle {\frac{{\hbar }^{5} k_i k'_j (k_k \!+\! k'_k)}{4 m \! \sqrt {\!A}}}} \!
\left( \!
\vk^2
\!+\!
\vk \!\cdot\! \vk'
\!+\!
\vk'^2 \!
\right)  , \\
\\[-10pt]
&
[v^{0,0(l)}_{\vk'''} , [v^{0,0(k)}_{\vk''}, [v^{0,0(j)}_{\vk'} , [v^{0,0(i)}_{\vk} , 
{T}_{0} (\rho )]]]]
\!=\!  
0  .
\EA
\right\}
\label{CRPiPiPiPiT0}  
\ea\\[-14pt]
By a procedure similar to the previous one,
we can determine the coefficients $\!{C_{\!n} (\!n \!\!\neq\!\! 0\!)}$
in
(\ref{T0rho})
and
then get an approximate form of ${T_{0}(\rho)}$
in terms of variables $\rho^{0,0}_{\vk}$ in the following form:\\[-22pt]
\ba
\!\!\!\!\!\!
\BA{lll}
{T}_{0} (\rho)
\!\!\!&=&\!\!\!\!\!
{C}_{0} 
\!+\!
{\displaystyle {\frac{{ \hbar }^{2}}{8m}}}
\sum_{\vk \ne 0} 
{\vk}^{2} \!
\rho^{0,0}_{\vk} \! \rho^{0,0}_{-\vk} \\
\\[-12pt]
\!\!\!&~~-&\!\!\!
{\displaystyle {\frac{{\hbar }^{2}}{24 m \! \sqrt {\!A}}}}
\sum_{\vp \ne 0, \vq \ne 0, \vp+\vq \ne 0} 
({\vp}^{2} \!+\! \vp \cdot \vq \!+\! {\vq}^{2})
\rho^{0,0}_{\vp} \rho^{0,0}_{\vq} \rho^{0,0}_{-\vp-\vq}
\!+\! 
O \! \left( \! {\displaystyle \frac{1}{A}} \! \right) \! . 
\EA
\label{T0rho2}  
\ea\\[-14pt]
With the aid of the underlying identities\\[-20pt]
\ba
\left.
\BA{c}
\sum_{\vp \ne 0, \vq \ne 0, \vp+\vq \ne 0} 
{\vp}^{2}
\rho^{0,0}_{\vp} \rho^{0,0}_{\vq} \rho^{0,0}_{-\vp-\vq} 
\!=\!
\sum_{\vp \ne 0, \vq \ne 0, \vp+\vq \ne 0} 
(\vp \!+\! \vq)^{2} \!
\rho^{0,0}_{\vp} \rho^{0,0}_{\vq} \rho^{0,0}_{-\vp-\vq}   ,\\
\\[-4pt]
\sum_{\vp \ne 0, \vq \ne 0, \vp+\vq \ne 0} 
{\vp}^{2}
\rho^{0,0}_{\vp} \rho^{0,0}_{\vq} \rho^{0,0}_{-\vp-\vq} 
\!=\! 
- 2 
\sum_{ p \ne 0, q \ne 0, p+q \ne 0} 
\vp \!\cdot\! \vq
\rho^{0,0}_{\vp} \rho^{0,0}_{\vq} \rho^{0,0}_{-\vp-\vq}  ,
\EA
\right\}
\label{identity}  
\ea\\[-12pt]
the lowest kinetic term  of $T$,
${T}_{0} (\rho)$
(\ref{T0rho2})
is rewritten as\\[-24pt]
\ba
\BA{c}
{T}_{0} (\rho)
\!=\! 
{C}_{0} 
\!+\!
{\displaystyle {\frac{{ \hbar }^{2}}{8m}}}
\sum_{\vk \ne 0} 
{\vk}^{2} \!
\rho^{0,0}_{\vk} \! \rho^{0,0}_{-\vk}
\!+\!
{\displaystyle {\frac{{\hbar }^{2}}{8 m \! \sqrt {\!A}}}}
\sum_{\vp \ne 0, \vq \ne 0, \vp+\vq \ne 0} 
\vp \!\cdot\! \vq
\rho^{0,0}_{\vp} \rho^{0,0}_{\vq} \rho^{0,0}_{-\vp-\vq} ~.
\EA
\label{T0rho3}  
\ea\\[-18pt]
From now on, we calculate the constant term $C_{0}$, 
the first term in the R.H.S. of
(\ref{T0rho2}).
Substituting
(\ref{T0rho2})  
into
(\ref{exactTPi}),
the constant term $C_{0}$ is computed up to the 
order of $\frac{1}{A}$:\\[-24pt]
\ba
\BA{lll}
{C}_{0} 
=
T
\!\!\!\!&-&\!\!\!\! 
{\displaystyle {\frac{{\hbar }^{2}}{8m}}} \!
\sum_{\vk \ne 0} 
{\vk}^{2} \!
\rho^{0,0}_{\vk} \rho^{0,0}_{-\vk} 
\!-\! 
{\displaystyle {\frac{1}{2m}}} \!
\sum_{\vk} \!
\vv^{0,0}_{\vk} \!\cdot\! \vv^{0,0}_{-\vk} 
\!+\!
{\displaystyle {\frac{1}{2 m \! \sqrt {\!A}}}} \!
\sum_{\vp+\vq \ne 0} \!
\rho^{0,0}_{\vp+\vq} \vv^{0,0}_{\vp} \!\cdot\! \vv^{0,0}_{\vq} \\
\\[-16pt]
\!\!\!\!&-&\!\!\!\!
{\displaystyle {\frac{{\hbar }^{2}}{8 m \! \sqrt {\!A}}}}
\sum_{\vp \ne 0, \vq \ne 0, \vp+\vq \ne 0} 
\vp  \!\cdot\! \vq
\rho^{0,0}_{\vp} \rho^{0,0}_{\vq} \rho^{0,0}_{-\vp-\vq} ~.
\EA
\label{C0phiexpansion}
\ea\\[-12pt]
Using
$\!\rho^{0,0}_{\vk}
\!\!\cong\!\!
\sum_{\tau_{z}} \!\!
\left( \!
\overline{\theta}a_{-\vk, \tau_{z}}
\!\!+\!\!
a^{\dag}_{\vk, \tau_{z}}\theta \!
\right)
$
and
$\vv^{0,0}_{\vk}
\!\!\cong\!\!
{\displaystyle \frac{\hbar \vk}{2}} \!
\sum_{\tau_{z}} \!\!
\left( \!
\overline{\theta}a_{\vk, \tau_{z}}
\!\!-\!\!
a^{\dag}_{-\vk, \tau_{z}}\theta \!
\right)
$,
we can calculate the third term in
(\ref{C0phiexpansion})
very similarly to the calculation of
(\ref{approxrhokvv}) 
and then reach to a result such as \\[-20pt]
\ba
\BA{l}
- 
{\displaystyle {\frac{1}{2m}}} \!
\sum_{\vk} \!
\vv^{0,0}_{\vk} \!\cdot\! \vv^{0,0}_{-\vk} 
\!=\!
{\displaystyle \frac{{\hbar }^{2}}{8 m}} \!
\sum_{\vk} \!
\vk^{2} \!
\sum_{\tau_{z}} \!\!
\left( 
\overline{\theta}a_{\vk, \tau_{z}} 
\!-\!
a^{\dag}_{-\vk, \tau_{z}}\theta 
\right) \!
\sum_{\tau'_{z}} \!\!
\left( 
\overline{\theta}a_{-\vk, \tau'_{z}} 
\!-\!
a^{\dag}_{\vk, \tau'_{z}}\theta 
\right) \\
\\[-10pt] 
\!=\!
{\displaystyle \frac{{\hbar }^{2}}{8 m}} \!
\sum_{\vk} \!
\vk^{2} \!
\left( 
\rho^{0,0}_{-\vk} 
\!-\!
2 \! \sum_{\tau_{z}} \! a^{\dag}_{-\vk, \tau_{z}}\theta 
\right) \!
\left( 
\rho^{0,0}_{\vk}
\!-\!
2 \! \sum_{\tau'_{z}} \! a^{\dag}_{\vk, \tau'_{z}}\theta 
\right) \\
\\[-10pt]
\!=\! 
{\displaystyle \frac{{\hbar }^{2}}{8 m}} \!
\sum_{\vk} \! 
\vk^{2}
\rho^{0,0}_{\vk} \! \rho^{0,0}_{-\vk}
\!-\! 
\theta \overline{\theta} 
\sum_{\vk} \!
{\displaystyle \frac{{\hbar }^{2} \vk^{2}}{2 m}} \! 
\sum_{ \tau_{z}, \tau'_{z}} \!
a^{\dag}_{\vk, \tau_{z}} \!
a_{\vk, \tau'_{z}}
\!+\!
\theta \overline{\theta}
{\displaystyle \frac{{\hbar }^{2}}{2 m}} \!
\sum_{\vk} \!
\vk^{2}
\!+\!
O \left( \! {\displaystyle \frac{1}{A}} \! \right)  .
\EA
\label{approxPiPi2}  
\ea\\[-16pt] 
 As for the forth and last terms in
(\ref{C0phiexpansion}),
due to the relations
$\theta \theta \!=\! 0$ and $\overline{\theta} ~\! \overline{\theta} \!=\! 0$,
we simply have\\[-22pt] 
\ba
\!\!\!\!\!\!\!\!
\BA{ll}
&~~
{\displaystyle {\frac{1}{2 m \! \sqrt {\!A}}}} \!\!
\sum_{\vp\!+\!\vq \ne 0} \!
\rho^{0,0}_{\vp\!+\!\vq} \vv^{0,0}_{\vp} \!\cdot\! \vv^{0,0}_{\vq} \\
\\[-10pt]
&\!\cong\! 
{\displaystyle {\frac{{\hbar }^{2}}{8 m \! \sqrt {\!A}}}} \!\!
\sum_{\vp\!+\!\vq \ne 0} \!
\vp  \!\cdot\! \vq \!
\sum_{\tau_{z}} \!\!
\left( \!
\overline{\theta} a_{-\vp\!-\!\vq, \tau_{z}} 
\!\!+\!\!
a^{\dag}_{\vp\!+\!\vq, \tau_{z}} \! \theta \!
\right) \!
\sum_{\tau'_{z}} \!\!
\left( \!
\overline{\theta} a_{\vp, \tau'_{z}} 
\!\!-\!\!
a^{\dag}_{-\vp, \tau'_{z}} \! \theta \!
\right) \!
\sum_{\tau''_{z}} \!\!
\left( \!
\overline{\theta} a_{\vq, \tau''_{z}} 
\!\!-\!\!
a^{\dag}_{-\vq, \tau''_{z}} \! \theta \!
\right)  \\
\\[-12pt]
& 
\!=\! 
0~\mbox{and also}
\EA
\label{rhopipi}
\ea\\[-14pt]
$
-
{\displaystyle {\frac{{\hbar }^{2}}{8 m \! \sqrt {\!A}}}}
\sum_{\vp \ne 0, \vq \ne 0, \vp+\vq \ne 0} 
\vp  \!\cdot\! \vq
\rho^{0,0}_{\vp} \rho^{0,0}_{\vq} \rho^{0,0}_{-\vp-\vq}
\!=\! 
0
$.
Substituting these into
(\ref{C0phiexpansion}), 
we have \\[-18pt]
\ba
\BA{l}
C_{0}
\!\cong\!
\left( 1 \!-\! \theta\overline{\theta} \right) \!
T
\!+\!
\theta\overline{\theta} 
{\displaystyle {\frac{{\hbar }^{2}}{2m}}} \!
\sum_{\vk} \! {\vk}^{2}
\!-\! 
\theta \overline{\theta} \! 
\sum_{\vk, \tau_{z} \ne \tau'_{z}} \!
{\displaystyle \frac{{\hbar }^{2} \vk^{2}}{2 m}} \! 
a^{\dag}_{\vk, \tau_{z}} \!
a_{\vk, \tau'_{z}} 
\!\cong\!
{\displaystyle {\frac{{\hbar }^{2}}{2 m}}} \!
\sum_{\vk} \! {\vk}^{2} .
\EA
\label{resultC0}
\ea\\[-10pt] 
Here we have used the relation
$\!\theta \overline{\theta} \!\!=\!\! 1\!$
and neglected the term
$\!
\sum_{\vk, \tau_{z} \ne \tau'_{z}} \!\!
 \frac{{\hbar }^{2} \vk^{2}}{2 m} \! 
a^{\dag}_{\vk, \tau_{z}} \!
a_{\vk, \tau'_{z}}
\!$
which does not exist
in the case of  the {\it isospin}-less Fermion system.
The result
(\ref{resultC0})
is not identical with the Sunakawa's result
\cite{SYN.62}
for a Bose system.
This is because
we have dealt with a Fermi system.
$\!\!$Then we get $\!$a$\!$ result which is considered
as a natural consequence for $\!$a$\!$ Fermi system.
It is surprising to see that the $C_0$
(\ref{resultC0})
coincides with the constant term in the
resultant ground state energy
given by the Tomonaga's method
\cite{Tomo.50}. 
Using
(\ref{exactTPi}), (\ref{T0rho2}) and (\ref{resultC0})
and separating the term
$\!C_{0} \!\!\cong\!\! \sum_{\vk} \! \frac{{\hbar }^{2}{\vk}^{2}}{2m}\!$
into two parts
$\!-\!\!\sum_{\vk} \! \frac{{\hbar }^{2}{\vk}^{2}}{4m}\!$
and
$\!
\frac{3}{2} \! \sum_{\vk} \! \frac{{\hbar }^{2}{\vk}^{2}}{2m} 
\!$,
we reach 
final goal of expressing the Hamiltonian $H$
(\ref{ExpressionforHamiltonian})
in terms of
$\rho^{0,0}_{\vk}$ and $\vv^{0,0}_{\vk}$
as follows:\\[-16pt]
\ba
\!\!\!\!
\BA{lll}
H 
\!\!\!\!&=&\!\!\!\!
-{\displaystyle \frac{A(A\!+\!2)}{8\Omega}} 
\nu (0) 
\!-\!
{\displaystyle \frac{A}{4\Omega}} \!
\sum_{\vk \ne 0} \!
\nu (\vk)
\!-\!
\sum_{\vk} 
{\displaystyle \frac{{\hbar }^{2}{\vk}^{2}}{4m}}
\!+\!
\underline{
{\displaystyle \frac{3}{2}} 
\sum_{\vk} 
{\displaystyle \frac{{\hbar }^{2}{\vk}^{2}}{2m}}
}  \\
\\[-14pt]
&&\!\!\!\!+
\sum_{\vk \ne 0} 
\left\{ 
{\displaystyle \frac{1}{2 m}}
\vv^{0,0}_{\vk}\!\cdot\! \vv^{0,0}_{-\vk}
\!+\!
\left( \!
{\displaystyle \frac{{\hbar }^{2}{\vk}^{2}}{8 m}} 
\!-\!
{\displaystyle \frac{A}{8\Omega}}
\nu (\vk) \!
\right) \!
\rho^{0, 0}_{\vk}
\rho^{0, 0}_{-\vk}
\right\} \\
\\[-18pt]
&&\!\!\!\!- 
{\displaystyle \frac{1}{2 m \! \sqrt {\!A}}} \!\!
\sum_{\!\vp \ne 0, \vq \ne 0, \vp\!+\!\vq \ne 0} \!
\vp \!\cdot\! \vq
\rho^{0,0}_{\!\vp+\vq} \vv^{0,0}_{\vp} \! \vv^{0,0}_{\vq} 
\!+\! 
{\displaystyle \frac{{\hbar }^{2}}{8 m \! \sqrt {\!A}}} \!\!
\sum_{\!\vp \ne 0, \vq \ne 0, \vp\!+\!\vq \ne 0} \!
\vp \!\cdot\! \vq
\rho^{0,0}_{\vp} \! \rho^{0,0}_{\vq} \! \rho^{0,0}_{\!-\vp-\vq} ,
\EA
\label{exactH}
\ea\\[-10pt]
in which the standard expression for the interaction $V$
in {\it isospin} $T\!=\!0$ nuclei 
is given by (2.9) and (2.10) in I.
In the case of 3D nuclei,
there appear volume $\Omega$ instead of length $L$.
$\!$Then we have the terms such as
$-{\displaystyle \frac{A}{4\Omega}}$
and
$-{\displaystyle \frac{A}{8\Omega}}$.
The expression
(\ref{exactH})
is just the Sunakawa's form up to the order of $\frac{1}{\sqrt A}$
\cite{SYN.62},
except the last term
$
\underline{
\frac{3}{2} \! \sum_{\vk} \! \frac{{\hbar }^{2}{\vk}^{2}}{2m}
} 
$
in the first line
of the R.H.S. of
(\ref{exactH}).
$\!\!$The second term
$\!
-\frac{A}{4\Omega}  
\nu (\vk)
\!$
also in the first line
is separated into two parts
$\!
\frac{A}{8\Omega}  
\nu (\vk)
\!$
and
$\!
-\frac{3A}{8\Omega}  
\nu (\vk)
$.
These differences arise due to the fact that
we deal with a {\it isospin} $T\!=\!0$ Fermi system
but not a Bose system.
At the present moment,
we discard the underlined term.
In
(\ref{exactH}),
the sum of some terms below
are considered as the lowest order Hamiltonian
$H_0$, \\[-22pt]
\ba
\!\!\!\!
\BA{c}
H_0
\!=\!
- {\displaystyle \frac{A \! \left(A \!+\!2 \right)}{8\Omega}} 
\nu(0) 
\!\!+\!\!
\sum_{ \! \vk \ne 0} \!\!
\left\{ \!
- \displaystyle
{{\frac{{\hbar }^{2} \! {\vk}^{2}}{4m}}
\!\!+\!\!
\frac{A}{8\Omega}
\nu(\vk)
\!\!+\!\!
{\displaystyle \frac{1}{2m}} \!
\vv^{0,0}_{\vk} \!\cdot\! \vv^{0,0}_{-\vk}
\!\!+\!\!
\left( \!\!
\frac{{\hbar }^{2} \! {\vk}^{2}}{8 m}
\!-\!
\frac{A}{8\Omega}
\nu(\vk) \!\!
\right) \!\!
\rho^{0,0}_{\vk} \!\! \rho^{0,0}_{-\vk}
} \!
\right\} \! .
\EA
\label{H0}
\ea\\[-16pt]
Now, let us introduce
the Boson annihilation and creation operators
defined as\\[-20pt]
\ba
\BA{c}
\alpha_{\vk}
\!\equiv\!
\sqrt{\!\displaystyle{\frac{ m E_{\vk}}{2{\hbar }^{2}{\vk}^{2}}}} 
\rho^{0,0}_{-\vk}
\!\!+\!\!
\displaystyle{\frac{1}{\sqrt{\!2 m {\vk}^{2} \! E_{\vk}}}}
\vk \!\cdot\! \vv^{0,0}_{\vk},~
\alpha^{\dag}_{\vk}
\!\equiv\!
\sqrt{\!\displaystyle{\frac{ m E_{\vk}}{2{\hbar }^{2} {\vk}^{2}}}} 
\rho^{0,0}_{\vk}
\!\!+\!\!
\displaystyle{\frac{1}{\sqrt{\!2 m {\vk}^{2} \! E_{\vk}}}}
\vk \!\cdot\! \vv^{0,0}_{-\vk},~\!
(\vk \!\ne\! 0) .
\EA
\label{Boson_ops}
\ea\\[-14pt]
Using
(\ref{Boson_ops})
and
(\ref{approxpirho}),
the collective variables
$\!\rho^{0,0}_{\vk}\!$ and $\!\vv^{0,0}_{\vk}\!$
are expressed as\\[-20pt]
\ba
\left.
\BA{ll}
&\rho^{0,0}_{\vk}
\!=\!
\sqrt{\!{\displaystyle \frac{{\hbar }^{2}{\vk}^{2}}{2 m E_{\vk}}}}
{\displaystyle \frac{1}{2}} \!
\left( 
\alpha_{-\vk}
\!+\!
\alpha^\dag_{\vk}
\right)
\!=\!
\sum_{\tau_{z}} \! 
\overline{\theta}a_{-\vk, \tau_{z}}
\!+\!
\sum_{\tau_{z}} \!
a^{\dag}_{\vk, \tau_{z}} \! \theta ,~\!
(\vk \!\ne\! 0) , \\
\\[-14pt]
&\vv^{0,0}_{\vk}
\!=\!
- i {\displaystyle \frac{\sqrt{2 m E_{\vk}}}{\vk \!\cdot\! \vk}}
{\displaystyle \frac{\vk}{2}} \!
\left( 
\alpha_{\vk}
\!-\!
\alpha^\dag_{-\vk}
\right)
\!=\!
{\displaystyle \frac{\hbar \vk}{2}} \!
\left( \!
\sum_{\tau_{z}} \!\! 
\overline{\theta}a_{\vk, \tau_{z}}
\!-\!
\sum_{\tau_{z}} \!\!
a^{\dag}_{-\vk, \tau_{z}} \! \theta 
\right) , ~\!
(\vk \!\ne\! 0) ,
\EA
\right\}
\label{rhoPi}
\ea\\[-10pt]
and substituting which into
(\ref{H0}),
the lowest order Hamiltonian
$\!H_{0}\!$
(\ref{H0})
is diagonalized as
\\[-20pt]
\ba
\BA{l}
H_0
\!=\!
E^G_0
\!+\!
\sum_{\vk \ne 0} E_{\vk}
\alpha^{\dag}_{\vk} \alpha_{\vk} , ~
E^G_0
\!\equiv\!
- {\displaystyle \frac{A \left(A \!+\! 2 \right)}{8\Omega}} 
\nu (0)
\!-\!
{\displaystyle \frac{1}{2}} 
\sum_{\vk \ne 0} 
\left( \!
E_{\vk}
\!-\!
{\displaystyle \frac{{\hbar }^{2} \! {\vk}^{2}}{2m}}
\!-\! 
{\displaystyle \frac{A}{2\Omega}} 
\nu (\vk) \!
\right) \! , \\
\\[-16pt]
E_{\vk}
\!\equiv\!
\sqrt{ \!
\left( 
\varepsilon_{\vk} 
\right)^2
\!-\!
{\displaystyle \frac{{\hbar }^{2} \! {\vk}^{2}}{m} 
\frac{A}{4\Omega}}
\nu(\vk)},  ~
\varepsilon_{\vk}
\!\equiv\!
{\displaystyle \frac{{\hbar }^{2} \! {\vk}^{2}}{2m}} ,~
(E_{\vk}\!:\mbox{Quasi-particle energy}) ,
\EA
\label{diagoH0}
\ea\\[-10pt]
where
we have used the commutation relation
$
[\vv^{0,0}_{\!\vk}, \rho^{0,0}_{\!\vk}]
\!=\! 
\hbar \vk 
$
given by
(\ref{modifiedCRs}).
In this sense,
the zero point energy of the collective mode
is included in the above diagonalization.
Since
$E_{\vk}$
is approximated as
$
\frac{{\hbar }^{2} \! {\vk}^{2}}{2m}
\!-\!
\frac{A}{4\Omega} 
\nu(\vk) ,
$
the term
$
\frac{1}{2}
( 
E_{\vk}
\!-\!
\frac{{\hbar }^{2} \! {\vk}^{2}}{2m}
\!-\!
\frac{A}{2\Omega} 
\nu(\vk) 
)
$
becomes
$
-
\frac{3A}{8\Omega} 
\nu(\vk) .
$
This is why we must separate 
$\!
-\frac{A}{4\Omega}  
\nu (\vk)
\!$
into 
$\!
\frac{A}{8\Omega}  
\nu (\vk)
\!$
and
$\!
-\frac{3A}{8\Omega}  
\nu (\vk)
$
but not arbitrary.
These facts lead us to $\!H_0$
(\ref{H0}).
$\!$The quantity $E^G_0$ in 
(\ref{diagoH0})
corresponds to the ground state energy.
Thus we have a Bogoliubov transformation
for$\!$ Boson-like operators
$\!\sum_{\tau_{z}} \!
\overline{\theta} a_{\vk, \tau_{z}}\!$
and
$\!\sum_{\tau_{z}} \!\!
a^{\dag}_{\vk, \tau_{z}} \!\! \theta$
as \\[-16pt]
\ba
\!\!\!\!\!\!
\BA{l}
\alpha_{\vk}
\!\!=\!\!
{\displaystyle
\frac{ \left( \! E_{\!\vk}\!\!+\!\!\varepsilon_{\!\vk} \! \right) \!\!
\sum_{\tau_{\!z}} \!\! \overline{\theta}a_{\vk, \tau_{\!z}}
\!\!\!+\!\!
\left( \! E_{\!\vk}\!\!-\!\!\varepsilon_{\vk} \! \right) \!\!
\sum_{\tau_{\!z}} \!\! a^{\dag}_{-\vk, \tau_{\!z}} \! \theta  }
{2 \sqrt{ \varepsilon_{\vk} E_{\vk}}}
} , ~
\alpha^{\dag}_{-\vk}
\!\!=\!\!
{\displaystyle
\frac{ \left( \! E_{\!\vk}\!\!-\!\!\varepsilon_{\!\vk} \! \right) \!\!
\sum_{\tau_{\!z}} \!\! \overline{\theta}a_{\vk, \tau_{\!z}}
\!\!\!+\!\!
\left( \! E_{\!\vk} \!\!+\!\!\varepsilon_{\!\vk} \! \right) \!\!
\sum_{\tau_{\!z}} \!\! a^{\dag}_{-\vk, \tau_{\!z}} \! \theta  }
{2 \sqrt{ \varepsilon_{\vk} E_{\vk}}}
} ,
\EA
\label{Bogolon_ops}
\ea\\[-10pt]
which is the same as the famous Bogoliubov transformation
for the usual Bosons
\cite{Bogoliubov.47}.
The diagonalization
(\ref{diagoH0})
has been
given similarly to the usual Bose system by Sunakawa
\cite{SYN2.62}.

%%%%%%%%%%%%%%%%%%%%%
%                                                                %
%  4   Discussions and further outlook      %
%                                                                %
%%%%%%%%%%%%%%%%%%%%%

\newpage

\def\thesection{\arabic{section}}
\setcounter{equation}{0}
\renewcommand{\theequation}{\arabic{section}.\arabic{equation}}
\section{Discussions and further outlook}

~~~
In the preceding sections,
we have proposed a velocity operator approach
to a 3D NP system.
After introducing collective variables,
the velocity operator approach
to the 3D NP system could be provided.
Particularly,
an interesting problem of describing
excitations occurring in nuclei,
$isospin~\!T \!=\! 0$ surface vibrations, 
may be possible to treat as an elementary exercise.
For this problem, for example,
see textbooks
\cite{RS.80,EG.87}.
By applying the velocity operator approach to such a problem,
an excellent description of the excitations
in $isospin~\!T \!=\! 0$ nuclei will be expected
to reproduce various correct behaviors including excited energies.
Because the present theory is constructed to take into account
important many-body correlations, which
have not been investigated sufficiently
for a long time 
in the historical ways for such a problem.
In this context,
it is said that
a new field of exploration of excitations
in a 3D Fermi system
may open
with aid of the velocity operator approach
whose new achievement may be appeared elsewhere.
By the way,
connection of the present theory with
the fluid dynamics was been mentioned briefly
by Sunakawa.
He transformed the quantum-fluid Hamiltonian
(\ref{exactH})
to the one in the configuration space
and obtained the classical-fluid Hamiltonian
for the case of the irrotational flow
\cite{SYN3.62,SYN2.62}.

There also exist $isospin~\!T \!\!=\!\! 1$ excitations
in nuclei.
As stressed in I,
the structures of the commutators among
$
\rho^{0,0}_{\vk}, \rho^{1,0}_{\vk},
\vg^{0,0}_{\vk} \! (\!=\! i \vk \pi^{0,0}_{\vk})
\!~\mbox{and}~\!
\vg^{1,0}_{\vk} \! (\!=\! i \vk \pi^{1,0}_{\vk})
$
have the twisted property
in the {\it isospin} space $(T,T_z)$.
Due to this fact,  
commutators
$[\vg^{1,0}_{\vk} , \rho^{1,0}_{\vk'}]\!$
and
$\![\vg^{1,0}_{\vk} , \vg^{1,0}_{\vk'}]$
are not closed.
The velocity operator
$\vv^{1,0}_{\vk}$
defined in the same way as
(\ref{modifiedv}) 
and density operator
 $\rho^{1,0}_{\vk}$
do not satisfy an important commutator
$
[\vv^{1,0}_{\vk} , \rho^{1,0}_{\vk'}]
\!=\!
\hbar \vk'  \delta_{\vk \vk'}
$.
Therefore,
the $\rho^{1,0}_{\vk}$ and $\vv^{1,0}_{\vk}$ are not
a suitable pair of collective operators for our object.
Then it turns out that
the isovector $T \!\!=\!\! 1$ surface vibrations
\cite{RS.80,EG.87}
can't be treated in the present approach.

As described in Section 3,
hitherto,
we have restricted Hilbert space to subspace
in which the {\bf vortex operator}
satisfies
$\mbox{rot}\vv^{0,0}(\vx)| \! >=0$,
i.e.,
$ 
[v^{0,0(i)}_{\vk} \! , v^{0,0(j)}_{\vk'}]
\!=\!
0
$.
While,
in the classical fluid dynamics,
the velocity field $\vv(\vx)$ is given as
$
\vv(\vx)
\!=\!
- \nabla \phi(\vx)
\!-\!
\lambda(\vx) \nabla \psi(\vx) ,
$
where $\phi(\vx)$ is the velocity potential and
$\lambda(\vx)$ and $\psi(\vx)$ are
Clebsch parameters
\cite{Clebsch.1857}.
This fact was already been pointed out by Sunakawa
\cite{SYN.62}.
As was suggested long time ago by Marumori $et~al$.
\cite{NagaTamaAmaMaru.58}
and Watanabe
\cite{Wata.56},
the {\bf internal rotational} motion, i.e., 
the vortex motion, however, may exist
also in nuclei.
So, we have something worthwhile in taking the vortex motion
into consideration.
In the very near future,
we will attempt at a description of {\bf rotational} velocity field
of a fluid in nuclei
through a Clebsch transformation.
Contrary to the above ways to the vortex motion in nuclei,
we should notice the paper in which
Holtzwarth and Sch\"{u}tte
have attempted at a derivation of fluid-dynamical equations of motion
which allow for velocity fields with vorticity,
starting from a time-dependent variational principle
for a many-fermion system.
They have derived an interesting relation between
the vorticity and the two-body correlations
\cite{HolSch.77}.
Standing on the above Clebsch viewpoint
and Ziman's
\cite{Ziman.53},
going from classical fluid dynamics to quantum fluid dynamics,
we will derive
a {\bf vortex} Hamiltonian of the fluid
in terms of roton operators.
The quantum fluid-dynamical approach
may be applied to a realistic nuclei.
Such an application to nuclei will provide
an excellent description of another kind of elementary energy excitation,
so-called the "{\bf vortex excitation}" occurred in nuclei
because the quantum fluid-dynamical manner may approach
various features of many-body effects,
which have  been discarded
in the traditional treatments
of the problem of rotational collective motion.
$\!\!$This work will be presented elsewhere in a forthcoming paper.

\newpage

%%%%%%%%%%%%%
%                                     %
%  Acknowledgements   %
%                                     %
%%%%%%%%%%%%%

\noindent
\centerline{\bf Acknowledgements}

\vspace{0.5cm}

One of the authors (S.N.) would like to
express his sincere thanks to
Professor Constan\c{c}a Provid\^{e}ncia for
kind and warm hospitality extended to him at
the Centro de F\'\i sica, Universidade de Coimbra.
This work was supported by FCT (Portugal) under the project
CERN/FP/83505/2008.

%%%%%%%%%%
%                           %
%   References      %
%                           %
%%%%%%%%%%

\end{document}